%% file: main.tex
\documentclass[journal]{IEEEtran}

\usepackage[T1]{fontenc}
\usepackage[utf8]{inputenc}
\usepackage{microtype}
\usepackage{amsmath,amssymb}
\usepackage{graphicx}
\usepackage{booktabs}
\usepackage{tabularx}
\usepackage{multirow}
\usepackage{makecell}
\usepackage{array}
\usepackage{xcolor}
\usepackage{cite}
\usepackage{url}
\usepackage{balance}
\usepackage{tikz}
\usetikzlibrary{arrows.meta,positioning,fit,shapes.geometric,shadows,calc,backgrounds}
\usepackage[most]{tcolorbox}
\usepackage{pifont}
\usepackage[hidelinks]{hyperref}
\usepackage[caption=false,font=footnotesize]{subfig}

\usepackage{booktabs}
\usepackage{tabularx}
\usepackage{array}
\usepackage{makecell}
\usepackage[table]{xcolor}

\newcolumntype{L}[1]{>{\raggedright\arraybackslash}p{#1}}

\definecolor{SMBlue}{HTML}{0B4F8A}
\definecolor{SMCyan}{HTML}{00A6C8}
\definecolor{SMPurple}{HTML}{6C4CCF}
\definecolor{SMGreen}{HTML}{22A06B}
\definecolor{SMOrange}{HTML}{E9792B}
\definecolor{SMDark}{HTML}{25313C}
\definecolor{SMLight}{HTML}{EEF5FA}
\definecolor{SMGray}{HTML}{667085}

\newtcolorbox{insightbox}[1]{
  colback=SMLight,
  colframe=SMBlue,
  boxrule=0.7pt,
  arc=1.2mm,
  left=1.2mm,right=1.2mm,top=1.0mm,bottom=1.0mm,
  fonttitle=\bfseries,
  title=#1
}

\newcolumntype{Y}{>{\raggedright\arraybackslash}X}
\newcolumntype{C}[1]{>{\centering\arraybackslash}p{#1}}

\renewcommand{\arraystretch}{1.08}

\title{StarMind: Challenges and Opportunities in Orbital Computing}

 \author{ \text{Zisheng Gong}\IEEEauthorrefmark{1}, \text{Zhaoyu Liu}\IEEEauthorrefmark{1}, \text{Houtianfu Wang}, \text{O. Tansel Baydas}, \text{Liu Cao},~\IEEEmembership{Member,~IEEE},  \text{Xia Xu}, \\ \text{Ozgur B. Akan},~\IEEEmembership{Fellow,~IEEE}

 \thanks{\IEEEauthorrefmark{1}Both authors contributed equally to this work. Zisheng Gong, Zhaoyu Liu and Liu Cao are with Electrical and Electronic Engineering Program of both City University of Hong Kong (Dongguan), Dongguan 523808, China and City University of Hong Kong, Hong Kong (e-mail:\{zisheng.gong, 72515198, liu.cao\}@cityu-dg.edu.cn).}
 \thanks{Xia Xu is with China Telecom Research Institute, Beijing 102209, China (e-mail:xiaxu@chinatelecom.cn).}
 \thanks{Houtianfu Wang, O. Tansel Baydas and Ozgur B. Akan are with the Centre for neXt Communications (CXC), Department of Engineering, University of Cambridge, CB3 0FA Cambridge, UK (e-mail:\{hw680, otb26, oba21\}@cam.ac.uk). Ozgur B. Akan is also with the Center for neXt Communications (CXC), Department of Electrical and Electronics Engineering, Ko\c{c}¸ University, 34450 Istanbul, T\"{u}rkiye (email:akan@ku.edu.tr).  (Corresponding author: Liu Cao)}

}

\begin{document}
\maketitle

\begin{abstract}

Space-based artificial intelligence (AI) is moving beyond isolated onboard processing toward constellation-scale computing over optical inter-satellite links (OISLs). Yet current infrastructures remain fragmented, making end-to-end (E2E) coordination across heterogeneous and multi-vendor systems difficult. This article examines the StarMind vision and develops a reference framework for distributed orbital computing.
It organizes the system into five layers and defines six standardizable interfaces that link mission intent to resource conditions and trustworthy result delivery. The framework also clarifies the operational constraints that determine whether orbital cooperation is worthwhile. In a representative synthetic aperture radar/interferometric SAR (SAR/InSAR) case study, the framework
reduces E2E latency to 41-min, compared with 61-min for localized onboard computing and 146-min for ground-based processing. This gain does not rely on faster processors in orbit. Instead, it comes from reducing contact waiting and urgent Earth downlink traffic through resource-aware task placement and gateway selection. Feasibility analysis shows that the proposed framework
is more effective when source datasets are large, output products are compact, and optical inter-satellite link capacity is sufficient. When these conditions are absent, conventional processing may remain preferable. Finally, we discuss research opportunities in joint task partitioning and cross-domain orchestration, hierarchical federated intelligence, and large model-enabled intelligence to advance an interoperable space-ground computing continuum.

\end{abstract}


\input{sections/01_introduction}

\input{sections/02_architecture}

\input{sections/03_standardization}
\input{sections/04_challenges}
\input{sections/05_opportunities}

\input{sections/06_roadmap}
\input{sections/07_conclusion}

\balance
\bibliographystyle{IEEEtran}
\bibliography{references}

\vspace{1mm}
\section*{Biographies}
\textbf{Zisheng Gong} received the M.S. degree in Electrical and Electronic Engineering from City University of Hong Kong, Hong Kong, in 2026, where he is currently a Research Assistant.

\textbf{Zhaoyu Liu} is currently pursuing the M.S. degree in Electrical and Electronic Engineering Program from City University of Hong Kong, Hong Kong.

\textbf{Houtianfu Wang} is currently pursuing the Ph.D. degree with the Centre for neXt Communications (CXC), Department of Engineering, University of Cambridge, CB3 0FA Cambridge, UK.

\textbf{O. Tansel Baydas} is currently pursuing the Ph.D. degree with the Centre for neXt Communications (CXC), Department of Engineering, University of Cambridge, CB3 0FA Cambridge, UK.

\textbf{Liu Cao} is currently an Assistant Professor at City University of Hong Kong (Dongguan), Dongguan 523808, China.


\textbf{Xia Xu} is a Tier-2 Chief Expert with the 6G Research Center, China Telecom Research Institute. He has made extensive contributions to international and domestic standardization in 3GPP and CCSA, and served as Vice Chair of 3GPP SA1 from 2019 to 2023. He currently serves as Vice Chair of CCSA TC12WG3 on satellite-terrestrial coordination. His research and standardization work has contributed to multiple 3GPP Release 19/20 specifications and industry standards.


\textbf{Ozgur B. Akan (Fellow, IEEE)} is currently a Professor of Electronic and Communication Engineering with the University of Cambridge, UK, where he leads the Centre for neXt Communications (CXC). He is also the Director of CXC at Ko\c{c} University, T\"{u}rkiye. His research interests include wireless, space, quantum, and molecular communications.



\end{document}

%% file: sections/01_introduction.tex
\section{Introduction}
\IEEEPARstart{S}{atellite} systems are evolving beyond their traditional roles as sensing and communication platforms. The increasing volumes of data generated by earth observation, astronomy, navigation, and space situational awareness are challenging the conventional workflow in which space-generated data are stored onboard, transmitted during limited ground contact windows, and subsequently processed in terrestrial data centers. Meanwhile, advances in onboard accelerators, optical inter-satellite links (OISL), orbital edge computing, and autonomous constellation management are enabling multiple satellites to jointly provide networking, storage, and computing services\cite{google_blog,10480332}. Motivated by these developments, a {\em StarMind} vision, proposed by SpaceX in 2026, represents a transition from isolated sensing and forwarding spacecraft toward an interconnected orbital intelligence fabric\cite{SpaceX_StarMind}.

Ground-based processing remains operationally mature and benefits from powerful, maintainable terrestrial computing infrastructures\cite{garcia2024advancements}. However, its data-to-information latency can be affected by limited ground contact opportunities and the downlink of large source datasets. Unlike conventional ground-based processing, StarMind enables selected tasks to be executed across distributed orbital computing and storage resources before compact results are delivered to the ground\cite{11588175}. The objective is not to replace terrestrial processing, but to reduce time-critical data movement and exploit constellation-level resources when in-orbit execution is advantageous.


Despite these advances, whether StarMind can deliver consistent system-level gains remains unresolved. Existing work on non-terrestrial networking, OISLs, onboard processing, satellite edge computing, and delay-tolerant networking has largely evolved along separate technical tracks, leaving the cross-layer orchestration of sensing, communication, storage, and computation immature. In addition, orbital workloads must operate under spacecraft-specific power, thermal, radiation, and maintainability constraints\cite{10480332,weston2025state}, while multi-vendor execution requires trustworthy and interoperable task, data, and result handling. The central question is therefore not whether all workloads should be moved into orbit, but under what conditions distributed in-orbit processing can outperform established ground-based processing.


Accordingly, this article presents a holistic picture of the envisioned StarMind architecture, covering its enabling technologies, communication protocols, system entities, service workflow, and standardizable interfaces. We identify the principal challenges associated with workload suitability, network--compute co-design, spacecraft constraints, on-orbit reliability, trustworthy operation, and sustainable deployment. To examine whether the envisioned architecture can provide measurable system-level benefits, we use synthetic aperture radar (SAR) and interferometric synthetic aperture radar (InSAR) processing as a representative data-intensive and time-sensitive case study \cite{garcia2024advancements}. Based on the resulting performance and feasibility boundaries, we further discuss three research opportunities for a space--ground computing continuum: joint task partitioning and cross-domain orchestration, hierarchical federated intelligence, and large-model-enabled space--ground intelligence.

%% file: sections/02_architecture.tex
\begin{figure*}[t]
    \centering
    \includegraphics[width=0.8\textwidth]{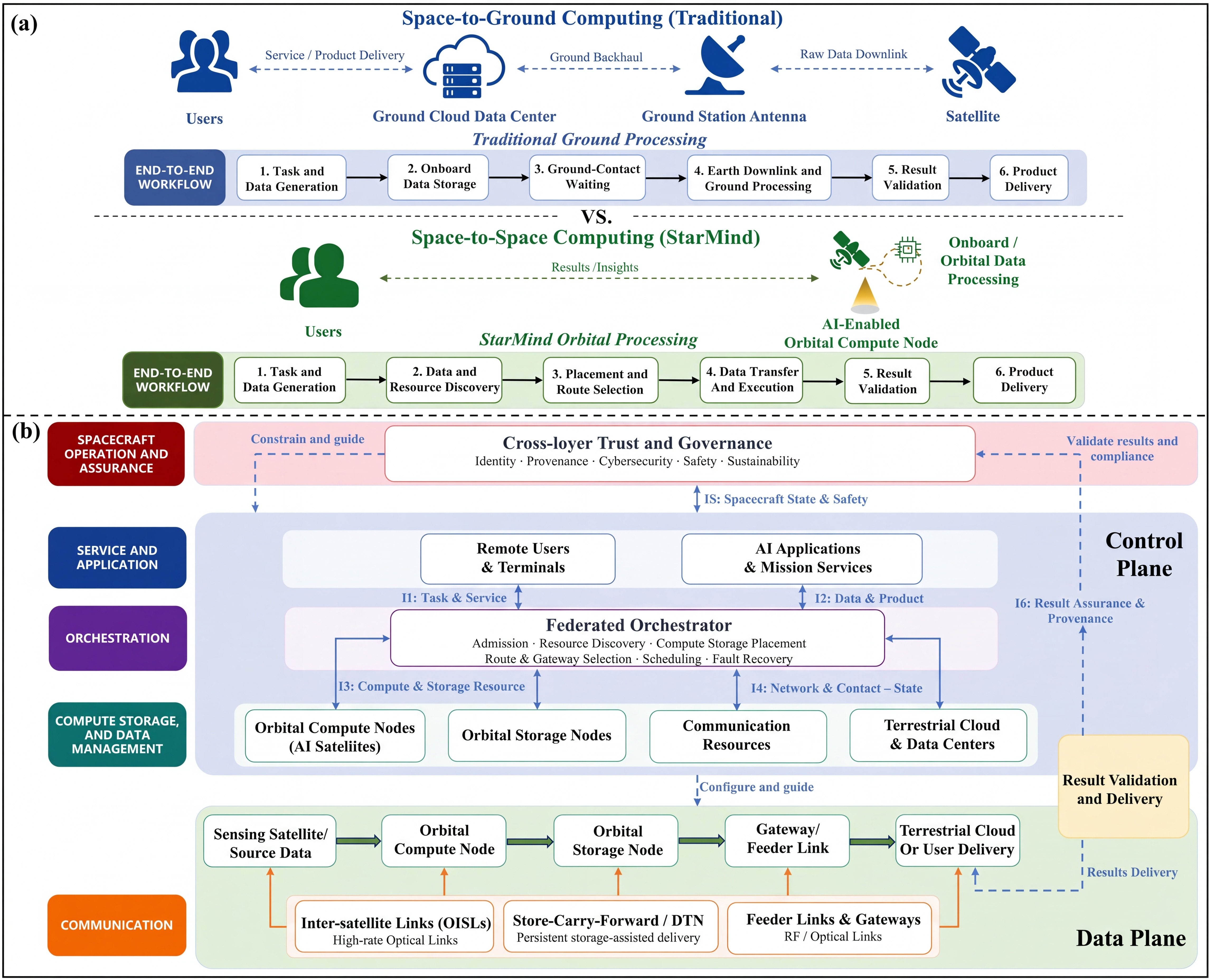}
    \caption{Computing Paradigm and Proposed Reference Architecture for StarMind: (a) Workflow Comparison between Traditional Space-to-ground and StarMind; (b) Five-layer Reference Architecture.}
    \vspace{-0.5cm}
    \label{fig:starmind_architecture}
\end{figure*}

\section{Architecture and System Framework}

\subsection{Distributed Orbital Computing Fabric}



Building on these developments, this article considers StarMind as a distributed orbital computing fabric that integrates sensing, communication, storage, computation, and autonomous orchestration across heterogeneous satellites. Unlike conventional onboard processing illustrated in Fig.~1(a), StarMind allows data and tasks to be assigned to suitable orbital nodes according to data locality, resource availability, link conditions, and service deadlines. Orbital storage can retain source, auxiliary, and intermediate data, while gateway satellites can exploit earlier ground contact opportunities for result delivery. Therefore, StarMind represents an evolution from isolated onboard processing toward constellation-level resource pooling, cross-satellite task placement, and resilient information delivery. Table~\ref{tab:paradigm_comparison} further positions StarMind relative to related satellite networking and computing paradigms. Rather than replacing SatCom/NTN, satellite edge computing, communication--computing integration, or DTN, StarMind builds on these foundations to enable constellation-level resource pooling and cross-layer orchestration.

\subsection{System Entities and Layered Architecture}
As Fig.~\ref{fig:starmind_architecture} shows, the envisioned StarMind architecture integrates heterogeneous orbital and terrestrial resources, including sensing satellites, orbital computing and storage nodes, gateways, and terrestrial infrastructures. Sensing satellites generate mission data, while computationally intensive or delay-sensitive tasks may be transferred to orbital compute nodes. Orbital storage nodes retain reusable data and execution states, and gateway satellites provide space--ground result delivery. Across the whole system, a federated orchestrator coordinates resource selection, placement, routing, and recovery across orbital and terrestrial resources. These entities describe where the sensing, computing, storage, and delivery functions reside, while an end-to-end (E2E) StarMind service further requires their coordinated operation. Building on the positioning in Table~I, we therefore propose a five-layer functional reference architecture in Fig.~1(b) to organize the heterogeneous capabilities envisioned by StarMind.

\begin{table*}[t]
\caption{Positioning of StarMind Relative to Related Satellite Networking
and Computing Paradigms}
\label{tab:paradigm_comparison}

\centering
\footnotesize
\setlength{\tabcolsep}{2.8pt}
\renewcommand{\arraystretch}{1.15}

\begin{tabularx}{\textwidth}{@{}
>{\bfseries\raggedright\arraybackslash}p{0.105\textwidth}
L{0.142\textwidth}
L{0.142\textwidth}
L{0.155\textwidth}
L{0.128\textwidth}
>{\raggedright\arraybackslash\columncolor{gray!7}}X
@{}}

\toprule
\makecell[l]{Comparison} &
\makecell[l]{\textbf{Traditional}\\\textbf{SatCom / NTN}} &
\makecell[l]{\textbf{Satellite Edge}\\\textbf{Computing}} &
\makecell[l]{\textbf{Communication--}\\\textbf{Computing Integration}} &
\makecell[l]{\textbf{DTN}} &
\cellcolor{gray!20}\makecell[l]{\textbf{StarMind}} \\
\midrule

\makecell[l]{Primary\\objective} &
Connectivity and data relay &
Processing close to the data source &
Joint communication--compute optimization &
Resilient delivery under disruption &
Distributed orbital computing fabric \\

\makecell[l]{Compute\\scope} &
Primarily terrestrial, with limited onboard processing &
Mainly local or edge nodes &
Multiple computing and networking resources &
Compute is not a native abstraction &
Heterogeneous orbital and terrestrial resources \\

\makecell[l]{Network\\role} &
Data transport &
Support for computation offloading &
Jointly optimized with computing &
Store-carry-forward transport &
OISL, DTN, and feeder connectivity jointly coordinated with task placement \\

\makecell[l]{Resource\\orchestration} &
Limited cross-domain orchestration &
Mostly node- or offloading-level scheduling &
Joint communication and computing resource management &
Routing- and contact-centric management &
Joint orchestration of computing, storage, networking, and spacecraft states \\

\makecell[l]{Key\\distinction} &
Mature satellite connectivity &
Onboard or edge processing &
Cross-resource optimization &
Intermittent and disruption-tolerant delivery &
Constellation-level resource pooling, cross-layer orchestration, and assurance \\

\bottomrule
\end{tabularx}

\vspace{-0.25cm}
\end{table*}

\begin{itemize}
     \item \textbf{Spacecraft Operation and Assurance Layer}
     exposes power, thermal, attitude, orbital, radiation, and fault states to the orchestrator. It also supports cybersecurity, provenance, operational safety, and trustworthy result delivery\cite{CCSDS_BundleProtocol_2015,ITU_RadioRegulations_2024,ISO_24113_2023}.

    \item \textbf{Service and Application Layer}
    translates mission or user intent into executable service requests. It specifies the requested information product, target region, deadline, priority, and quality or security requirements\cite{SpaceX_StarMind}.

    \item \textbf{Orchestration Layer}
    maps service requests onto available orbital and terrestrial resources. It coordinates task admission, resource placement, E2E scheduling, and fault recovery according to service requirements and system states\cite{3GPP_TS28533}.

    \item \textbf{Compute, Storage, and Data Management Layer}
    provides heterogeneous processors, runtime environments, and persistent storage for distributed task execution. It also manages data catalogs, software or model versions, intermediate products, task states, and checkpoints\cite{melega2025orbit}.

    \item \textbf{Communication Layer}
    supports space--space and space--ground data transfer through OISLs, feeder links, routing, contact planning, and disruption-tolerant networking. It carries source data, intermediate states, and final products among sensing, computing, storage, and gateway nodes\cite{3GPP_TR38821_2023,CCSDS_BundleProtocol_2015}.
\end{itemize}

\begin{figure*}[t]
\centering
\includegraphics[width=.8\textwidth]{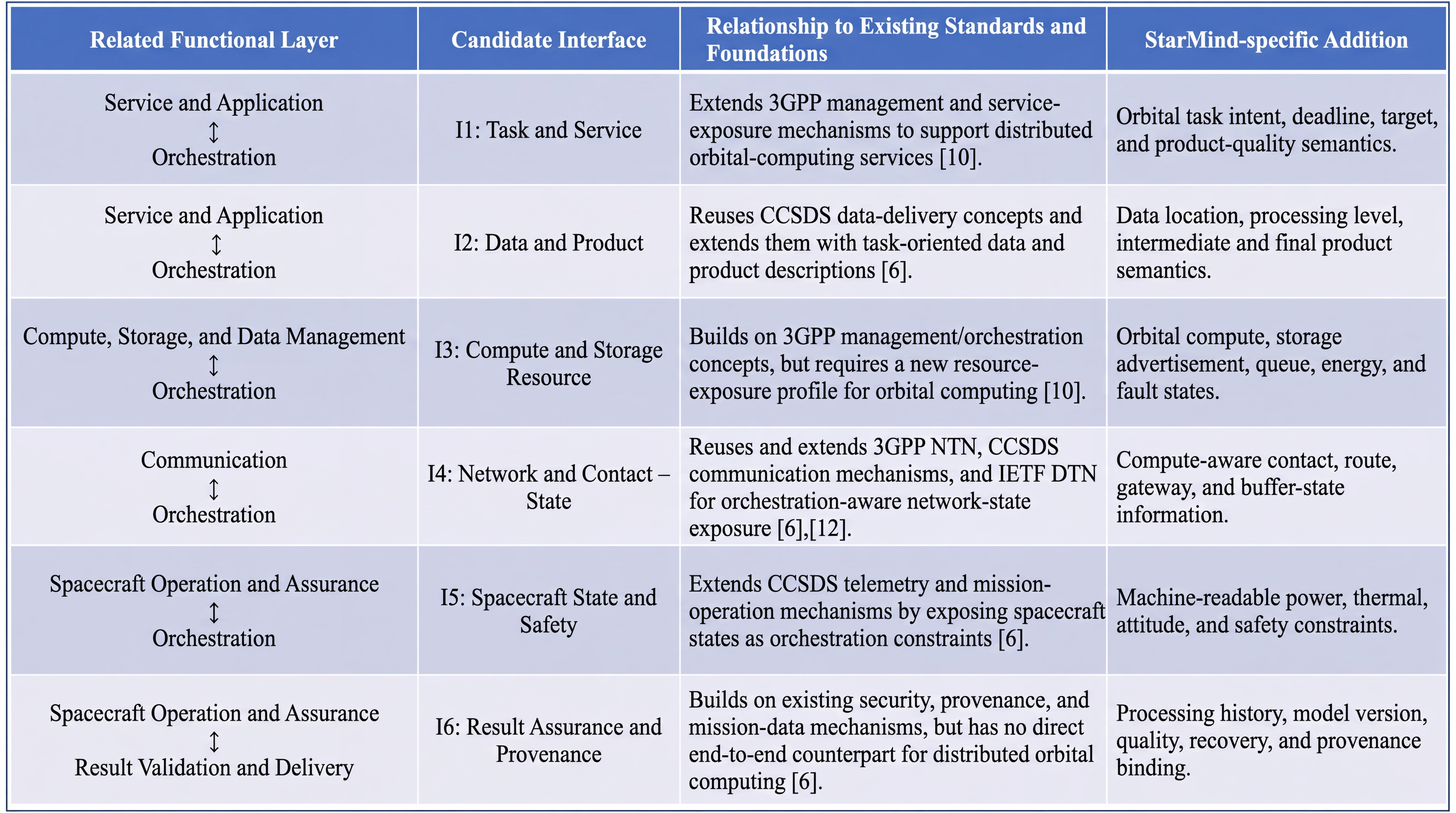}
\vspace{-0.3cm}
\caption{Standardization Mapping of Candidate StarMind Interfaces to Existing Foundations.}
\vspace{-0.5cm}
\label{fig:stack}
\end{figure*}

\vspace{-0.5cm}
\subsection{E2E Workflow and Standardizable Interfaces}
The StarMind service is realized through coordinated control-plane decision-making and data-plane execution, as illustrated in Fig.~\ref{fig:starmind_architecture}(b). A sensing satellite, mission control center, application, or ground user first specifies the requested product, target region, deadline, priority, and quality requirements. The control plane then collects the relevant service requirements, data and resource states, network conditions, and spacecraft operating constraints, based on which the federated orchestrator determines task placement, routing, and execution scheduling. These decisions configure and guide the data plane, where source data, intermediate products, and task states are transferred and processed across orbital compute and storage nodes, OISLs, DTN, feeder links, and terrestrial infrastructures. After execution, the resulting product is validated and delivered to the requesting user or terrestrial data center, while assurance and provenance information support trustworthy service completion.








Coordinating this workflow across heterogeneous and independently operated systems requires common information exchanges between the control-plane functions and the underlying execution resources. Accordingly, we propose six
candidate standardizable interfaces as logical information exchanges,
rather than replacement protocol stacks. These interfaces are intended
to reuse existing 3GPP, CCSDS, and IETF DTN mechanisms wherever
applicable, while introducing only the additional task, resource,
network, spacecraft-state, and assurance semantics required for
distributed orbital computing\cite{3GPP_TS28533,CCSDS_BundleProtocol_2015}:

\begin{itemize}
    \item \textbf{Interface 1 (I1): Task and Service Interface}
    specifies the task type, requested product, deadline, priority, target region, and quality or security requirements.

    \item \textbf{Interface 2 (I2): Data and Product Interface}
    describes data identity, format, processing level, size, location, calibration metadata, and product quality information.

    \item \textbf{Interface 3 (I3): Compute and Storage Resource Interface}
    advertises processor type, available computing capacity, memory, storage, queue state, fault status, and energy budget.

    \item \textbf{Interface 4 (I4): Network and Contact-State Interface}
    provides link capacity, OISL availability, contact windows, route 
    reliability, buffer state, and gateway visibility.

    \item \textbf{Interface 5 (I5): Spacecraft State and Safety Interface}
    exposes power, thermal, attitude, pointing, orbital, maneuver, and fault constraints to the orchestrator.

    \item \textbf{Interface 6 (I6): Result Assurance and Provenance Interface}
    records input-data identity, processing history, software or model version, quality indicators, integrity signatures, and recovery events.
\end{itemize}

The above interfaces form the logical coordination fabric between the control plane and the data plane. I1 and I2 convey service intent and data/product information to the orchestrator; I3, I4, and I5 expose the compute and storage resources, network/contact conditions, and spacecraft operating constraints required for placement, routing, and scheduling. The resulting control decisions then configure and guide data-plane execution, in which source data, intermediate products, task states, and results are physically transferred and processed. Finally, I6 associates the validated result with assurance and provenance information, closing the loop between execution and trustworthy service delivery. Fig.~\ref{fig:stack} maps these interfaces to existing standards and highlights where StarMind reuses established mechanisms or requires additional orbital computing semantics. The mapping emphasizes extension and integration of the existing 3rd Generation Partnership Project (3GPP), Consultative Committee for Space Data Systems (CCSDS), and the Internet Engineering Task Force (IETF) DTN foundations \cite{3GPP_TR38821_2023}.

%% file: sections/04_challenges.tex
\section{Key Challenges}

Although onboard processing, OISLs, and satellite networking have been individually demonstrated, integrating them into a general-purpose orbital computing fabric introduces several system-level challenges. The core difficulty is not simply placing more processors in space, but ensuring that the available computing, communication, storage, and spacecraft resources can be effectively utilized to deliver useful information products. The major challenges are discussed below.

\subsubsection{Challenge 1 - Benefit-Aware Workload Selection}

In-orbit processing is most beneficial when source datasets are large, output products are compact, and ground contact or earth downlink delays dominate. Onboard cloud filtering and information extraction  are representative examples because uninformative observations can be discarded before downlink\cite{melega2025orbit}. In contrast, workloads that generate large 
intermediate products, require frequent synchronization, or ultimately demand the complete raw data on the ground may obtain limited benefit from orbital processing. StarMind should therefore be evaluated by E2E product latency and traffic reduction, including orbital transmission, computation, queueing, validation, and system-availability overheads, rather than peak computing capacity alone.


\subsubsection{Challenge 2  -  Joint Network-Compute-Storage Orchestration}

Orbital computing architecture determines the balance between computation and communication. Local execution minimizes OISL traffic but may underutilize distributed resources, whereas clustered or constellation-wide execution increases resource pooling at the cost of transmission, synchronization, and recovery overhead. StarMind therefore requires joint task placement and network control: distributing a workload is beneficial only when execution-time savings exceed the additional coordination cost.

\subsubsection{Challenge 3  -  Spacecraft-Constrained Reliable Execution}

Orbital compute scheduling must account for shared power, thermal, communication, attitude, and payload constraints. Available capacity varies with illumination, battery state, mission activity, and heat-rejection capability, requiring power capping, thermal-aware placement, throttling, or migration. Radiation and limited maintainability further motivate error detection, checkpointing, replication, and constellation level recovery, all of which consume the same resources they protect.

\begin{figure*}[!t]
    \centering
    \subfloat[E2E Latency Breakdown.]{
        \includegraphics[width=.315\textwidth]{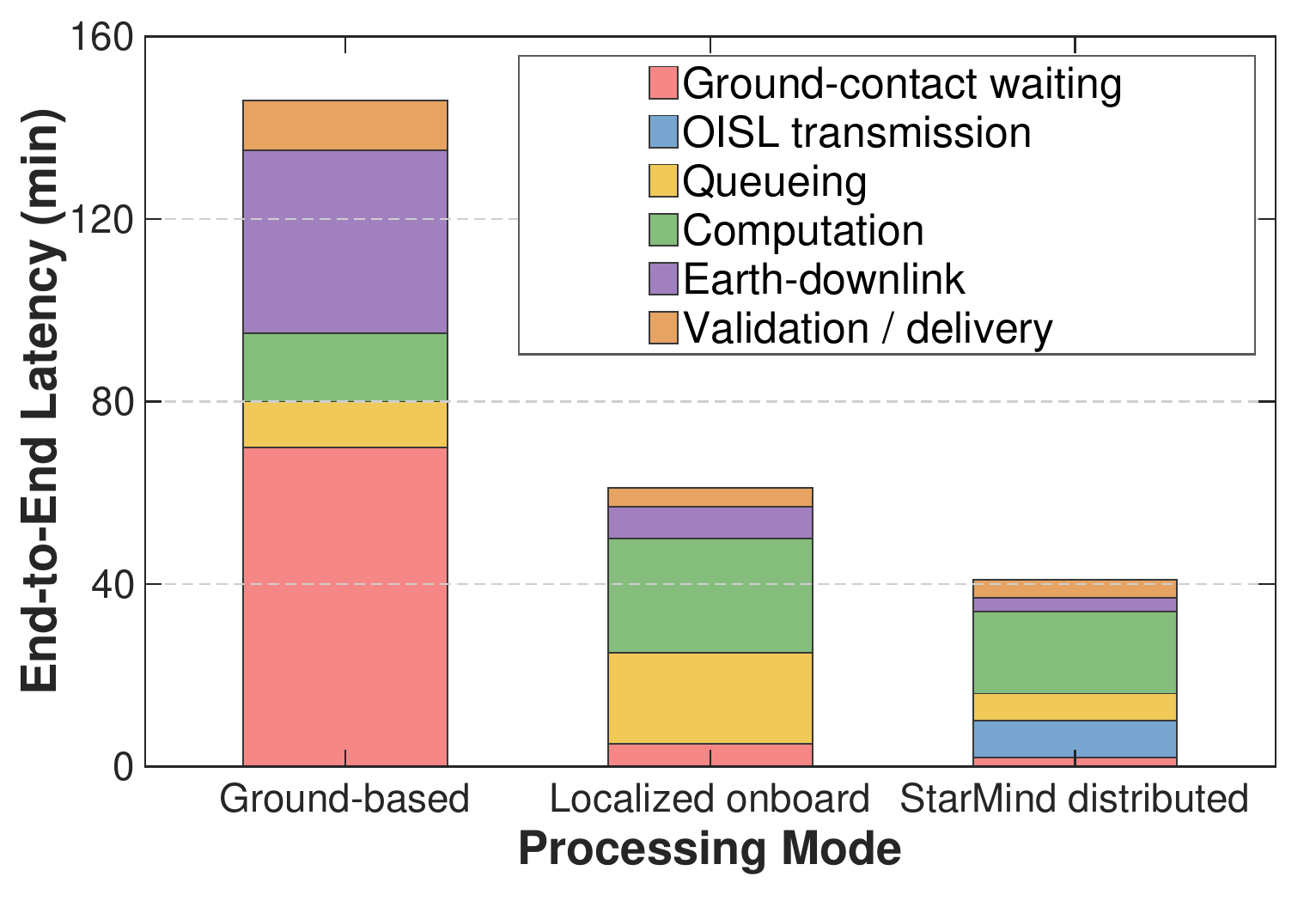}
        \label{fig:latency_breakdown}
    }
    \hfill
    \subfloat[E2E Latency CDF.]{
        \includegraphics[width=.315\textwidth]{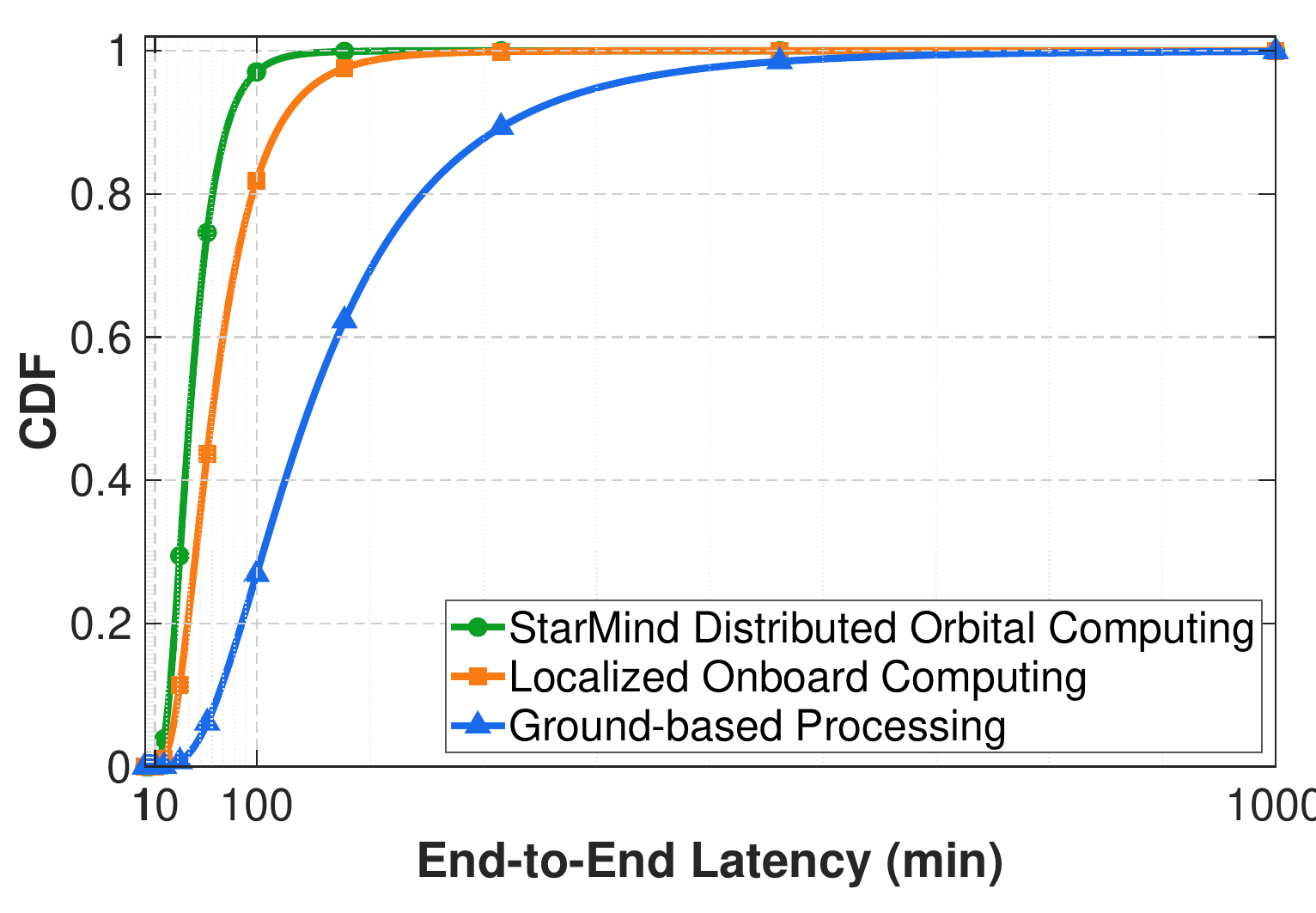}
        \label{fig:latency_cdf}
    }
    \hfill
    \subfloat[Earth Downlink Volume.]{
        \includegraphics[width=.315\textwidth]{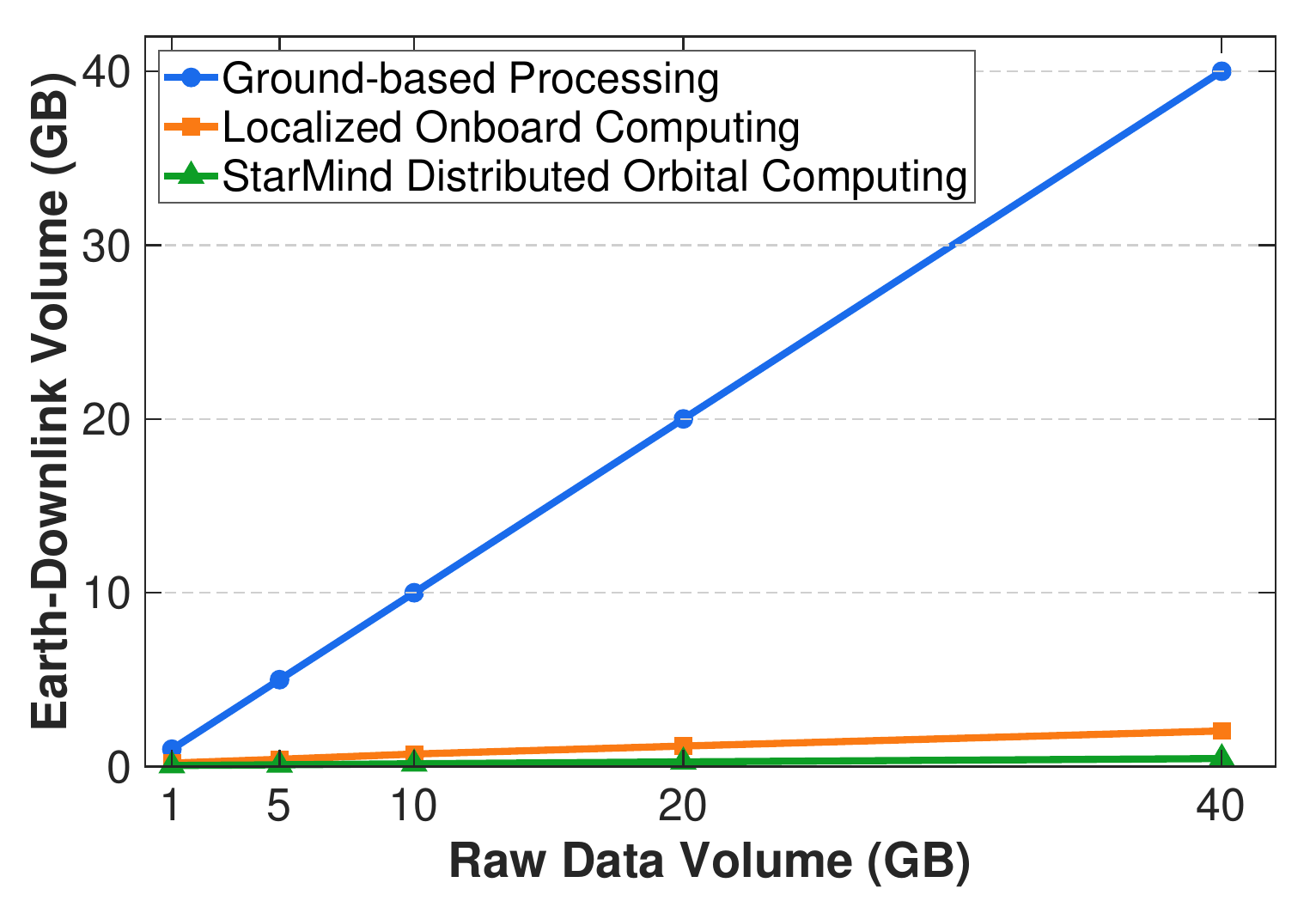}
        \label{fig:downlink_volume}
    }

    \caption{Case Study Comparison among Ground-based Processing, Localized Onboard Computing, and StarMind Distributed Orbital Computing.}
    \label{fig:case_study_performance}
    \vspace{-0.8cm}
\end{figure*}


\begin{figure*}[!t]
    \centering
    \subfloat[Impact of OISL Bandwidth and Ratio of Raw Data Size to Product Size.]{
        \includegraphics[width=.48\textwidth]{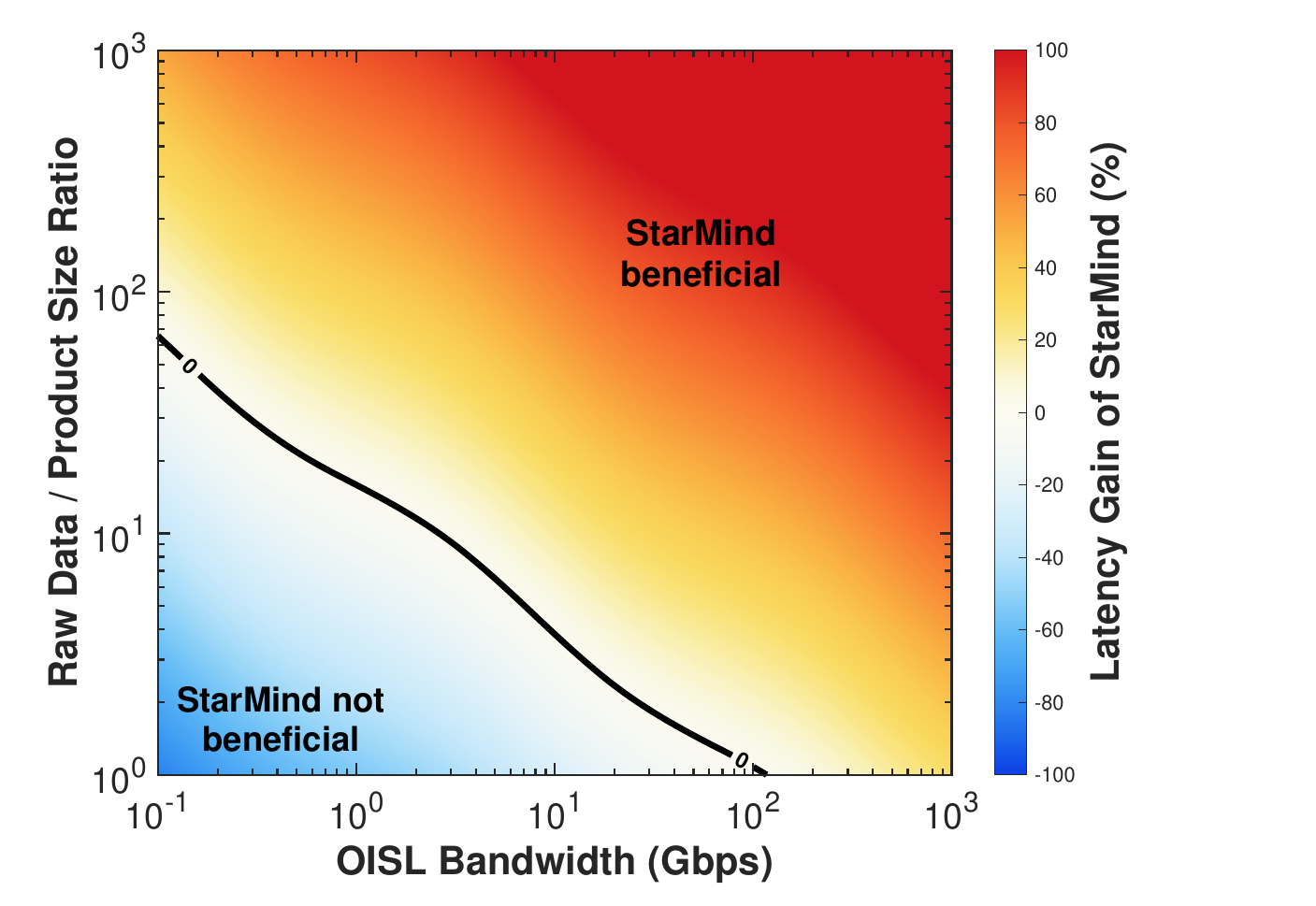}
        \label{fig:feasibility_oisl}
    }
    \hfill
    \subfloat[Impact of Ground Contact Waiting Time and Raw Data Volume.]{
        \includegraphics[width=.48\textwidth]{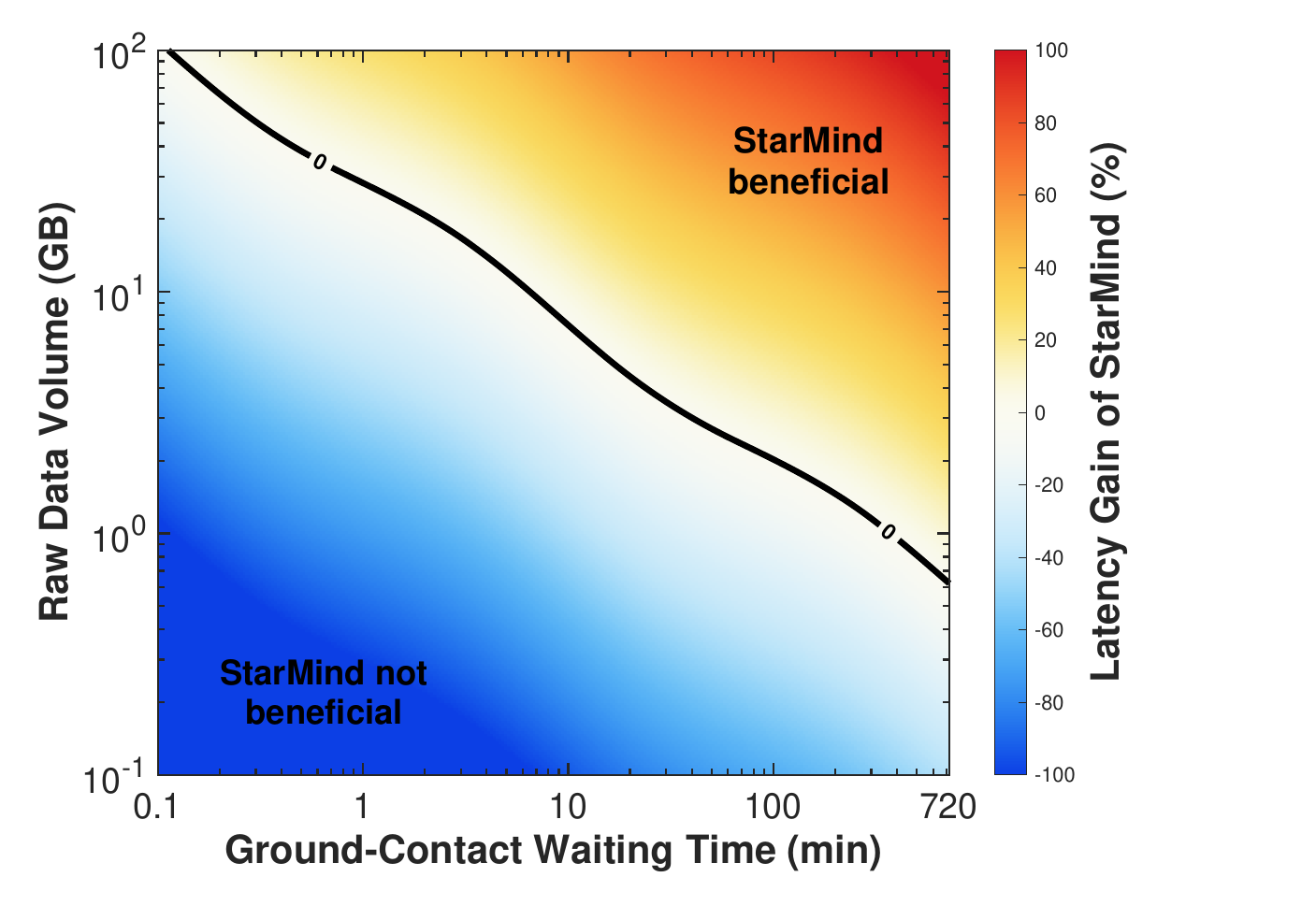}
        \label{fig:feasibility_waiting}
    }
    \caption{Feasibility Regions of StarMind Distributed Orbital Computing. }
    \label{fig:starmind_feasibility}
       \vspace{-0.5cm}
\end{figure*}

\subsubsection{Challenge 4  -  Economics, Trust, and Sustainability}

The economic feasibility of orbital computing should be evaluated by the cost of delivering a valid information product rather than peak computing power, considering launch and manufacturing costs, utilization, lifetime, replacement, OISL/storage occupancy, fault-tolerance overhead, and avoided Earth downlink traffic. It is viable only when latency and data-movement gains offset deployment costs and limited hardware utilization.

Scalable operation also requires interoperable and trustworthy cross-vendor execution, supported by common task, data, resource, spacecraft-state, and provenance descriptions. Large constellations must further coordinate spectrum and orbital resources. Although OISLs reduce radio-frequency dependence, telemetry, command, access, and feeder links remain subject to coordinated spectrum use and applicable ITU procedures \cite{ITU_RadioRegulations_2024}. Larger fleets also intensify collision avoidance, disposal, re-entry, and debris mitigation requirements under ISO~24113 \cite{ISO_24113_2023}. StarMind must therefore balance performance, cost, interoperability, and sustainability.

\section{Case Study: In-Orbit versus Ground-Based SAR/InSAR Processing}



\begin{table*}[!t]
\centering
\caption{Representative Research Opportunities for Space--Ground Collaborative Computing}
\vspace{-0.3cm}
\label{tab:sg_opportunities}
\renewcommand{\arraystretch}{1.18}
\setlength{\tabcolsep}{4pt}
\small
\begin{tabularx}{\textwidth}{
    >{\raggedright\arraybackslash}p{0.31\textwidth}
    >{\raggedright\arraybackslash}p{0.2\textwidth}
    >{\raggedright\arraybackslash}X}
\toprule
\textbf{Research Opportunities} &
\textbf{Required interfaces} &
\textbf{Recommended metrics} \\
\midrule

Joint task partitioning and cross-domain orchestration &
I1–I5 & 
Latency, deadline miss, traffic, energy, thermal violations \\

Hierarchical federated intelligence across space and ground &
I2–I6 &
Model accuracy, convergence time, model freshness, communication overhead, robustness \\

Large-model-enabled space--ground intelligence &
I1–I6 & 
Task success, uncertainty calibration, verification pass rate, latency, memory and energy \\
\bottomrule
\end{tabularx}
\vspace{-0.5cm}
\end{table*}

We conduct an illustrative system-level evaluation to examine when distributed in-orbit processing can deliver a rapid SAR/InSAR product earlier than conventional alternatives. The requested product may be an interferogram, coherence map, deformation estimate, region of interest, or event alert. Three execution modes are considered:
\begin{itemize}
  \item \textbf{Ground-based processing (GND):} The sensing satellite waits for ground contact, downlinks the raw or partially processed data, and leaves the complete workflow to a terrestrial server.

  \item \textbf{Localized onboard computing (LOCAL):} The sensing satellite processes the data locally and downlinks the compact product at the next contact, without using computing resources on other satellites.

  \item \textbf{StarMind distributed orbital computing (SM):} Data are transferred via OISLs to a selected orbital compute node, and the product may be relayed through an earlier-visible gateway based on resource load and predicted connectivity.
\end{itemize}

 \vspace{-0.5cm}
\subsection{Evaluation Setup and Latency Model}



All modes use the same source data, processing objective, output format, and quality requirement. The timer starts when data acquisition is completed and ends when the requested product has passed its quality check and is available to the ground user. Reference observations and reusable auxiliary data are assumed to have been pre-positioned at the required storage nodes. The ground server is more computationally capable than an individual orbital node, while a dedicated StarMind computing satellite may be faster than the processor on the sensing satellite. 


The E2E latency is decomposed into contact-window waiting, data transmission, compute queueing, task execution, and product validation and delivery:



\begin{equation}
T_{\rm GND}=\frac{D_{\rm raw}}{R_{\rm SG}^{\rm GND}}+T_{\rm wait}^{\rm raw}+T_q^{g}+T_c^{g}+T_{\rm vd},
\end{equation}

\begin{equation}
T_{\rm LOCAL}=\frac{D_{\rm raw}}{\rho R_{\rm SG}^{\rm LOCAL}}+T_{\rm wait}^{\rm prod}+T_q^{l}+T_c^{l}+T_{\rm vd},
\end{equation}

\begin{equation}
T_{\rm SM}=\frac{D_{\rm raw}}{\rho R_{\rm SG}^{\rm SM}}+\frac{D_{\rm raw}}{R_{\rm OISL}}+T_{\rm wait}^{\rm gw}+T_q^{o}+T_c^{o}+T_{\rm vd},
\end{equation}
where $D_{\rm raw}$ denote the source data, and $\rho$ is the raw-data-to-product size ratio. $R_{\rm SG}^{\rm GND}$, $R_{\rm SG}^{\rm LOCAL}$, and $R_{\rm SG}^{\rm SM}$ denote the effective space-ground transmission rates available to GND, LOCAL, and SM, respectively, while $R_{\rm OISL}$ is the effective OISL rate between the sensing satellite and the selected orbital computing node. $T_{\rm wait}^{\rm raw}$ denotes the waiting time from data acquisition to a sensing satellite contact capable of transmitting the source data, $T_{\rm wait}^{\rm prod}$ is the waiting time from completion of local processing to a product-capable contact, and $T_{\rm wait}^{\rm gw}$ is the waiting time from completion of orbital processing to a usable contact of the selected gateway. These waiting times are derived from the same underlying contact timeline but can differ because the three modes become transmission-ready at different times and may use different contact opportunities. $T_q^{g}$, $T_q^{l}$, and $T_q^{o}$ denote queueing delays at the terrestrial server, sensing satellite, and orbital computing node, respectively, while $T_c^{g}$, $T_c^{l}$, and $T_c^{o}$ are their corresponding computation times. $T_{\rm vd}$ covers validation, provenance generation, and final delivery.

\vspace{-0.3cm}
\subsection{Results and Insights}

Fig.~\ref{fig:case_study_performance} compares the three execution modes under the above baseline. As shown in Fig.~\ref{fig:case_study_performance}\subref{fig:latency_breakdown}, GND has the shortest computation time but incurs a 70-min wait for a raw data capable contact and a 40-min source data downlink. LOCAL avoids raw data transmission but incurs longer onboard queueing and computation, yielding 61-min in total. SM introduces an 8-min OISL transfer, but resource-aware compute node and gateway selection reduces its queueing and delivery waiting, yielding 41-min. The resulting GND latency is 146-min. To examine latency variability without prescribing separate output distributions for the three modes, Fig.~\ref{fig:case_study_performance}\subref{fig:latency_cdf} reports empirical CDFs from $10^4$ Monte Carlo realizations. Each realization perturbs the contact opportunities, queueing loads, computation times, and effective link rates around the baseline, while all three modes are evaluated under the same sampled task and contact realization. The resulting distributions illustrate the lower typical and tail latencies of SM under the assumed variability. Fig.~\ref{fig:case_study_performance}\subref{fig:downlink_volume} further varies the source data volume from 1 to 40 GB and shows that GND's urgent Earth downlink grows directly with source data size, whereas LOCAL and SM mainly transmit compact products. Thus, the benefit of SM is earlier information delivery and reduced urgent Earth downlink rather than elimination of communication.

We next evaluate the operating region of distributed orbital computing using the same latency model. The latency gain of SM is defined as the percentage reduction in E2E latency relative to the better conventional alternative, i.e., the one with the smaller latency between GND and LOCAL. A positive gain therefore indicates that SM outperforms both conventional alternatives, while zero gain defines the break-even boundary.
In Fig.~\ref{fig:starmind_feasibility}\subref{fig:feasibility_oisl}, $D_{\rm raw}=40$ GB and the baseline contact and computing parameters are fixed, while the OISL bandwidth is swept from 0.1 to 1000 Gbps and $\rho$ from 1 to 1000. Higher OISL capacity and stronger data reduction enlarge the beneficial region because they reduce task migration and result delivery overhead. Fig.~\ref{fig:starmind_feasibility}\subref{fig:feasibility_waiting} fixes $\rho=10$ and $R_{\rm OISL}=0.667$ Gbps and varies the waiting time to the next raw data capable contact from 0.1 to 720-min together with $D_{\rm raw}$ from 0.1 to 100 GB. The results indicate that higher OISL capacity and stronger data reduction enlarge the beneficial region of SM, whereas slow OISLs and limited data reduction make migration overhead dominant. Likewise, small datasets with imminent ground contact may favor ground-based processing, while large datasets and long contact waits favor distributed orbital processing. The zero gain contours therefore highlight that orbital execution should be selected adaptively according to workload and connectivity rather than applied as a fixed processing mode. 

%% file: sections/05_opportunities.tex
\section{Research Opportunities of Space--Ground Collaborative Computing}


Distributed in-orbit and ground-based processing should be integrated into a space--ground computing continuum, where tasks, data, and models are placed according to connectivity, computing availability, service deadlines, and product quality requirements. Table~\ref{tab:sg_opportunities} summarizes three representative research opportunities with the required interfaces and recommended metrics.

\vspace{-0.3cm}
\subsection{Joint Task Partitioning and Cross-Domain Orchestration}

A space--ground workflow can partition tasks across sensing satellites, orbital computing nodes, and terrestrial servers. Lightweight preprocessing or rapid product generation may be performed near the data source, while compute-intensive stages are assigned to more capable nodes. Processor capability alone, however, does not determine the best placement: a lightly loaded satellite may require a long multi-hop OISL path, whereas a powerful ground server may be unavailable until the next contact window. Parallel execution can also shorten computation while increasing intermediate data transfer and synchronization overhead. Orchestration must therefore balance computing, communication, storage, energy, and service deadlines. A hierarchical approach can coordinate these decisions across timescales. Slow-timescale planning provisions resources using predicted demand and connectivity, whereas fast-timescale scheduling adapts execution to current workloads and system states \cite{zhou2025mission}. Common descriptions of tasks, data, software, and resources are also needed to support collaboration among independently operated orbital and terrestrial platforms.

\vspace{-0.3cm}
\subsection{Hierarchical Federated Intelligence across Space and Ground}

Space--ground collaboration can support distributed model training without repeatedly transferring complete raw datasets. Satellites perform local training and exchange model updates or compact knowledge representations, while terrestrial infrastructures provide complementary data and scalable computing. Under intermittent connectivity, satellites in the same orbital region may first exchange updates through OISLs and aggregate them at an orbital node. The intermediate model can then be forwarded through a gateway and combined with terrestrial updates. Compared with direct satellite-to-ground aggregation, this hierarchy reduces repeated space--ground exchanges and exploits orbital communication locality \cite{shi2024satellite}. Orbital and terrestrial participants nevertheless differ in computing capability, memory, energy, contact duration, and local data distribution, leading to heterogeneous delays and model freshness. Future work should jointly address participant and aggregator selection, communication efficiency, freshness, and trustworthy aggregation under these constraints.

\subsection{Large Model-Enabled Space-Ground Intelligence}

Large language models, vision foundation, and multimodal models could extend StarMind from task-specific processing to more general intelligence services. They may translate natural-language mission requests into processing graphs, identify required sensing and auxiliary data, select analysis tools, and generate human-readable products. For Earth observation, multimodal models may combine optical imagery, SAR products, telemetry, historical observations, and textual knowledge for disaster interpretation, anomaly analysis, and rapid reporting. Because complete large models may exceed spacecraft memory and computing budgets, a practical architecture should combine compact orbital models with larger models on orbital computing satellites or terrestrial data centers. Sensing satellites may perform lightweight encoding, event detection, or small-model inference, while selected features, semantic tokens, or uncertain samples are transferred for split inference, expert selection, retrieval, or federated fine-tuning \cite{zhu2025satellite}.

Large models could also assist orchestration by interpreting service intent, proposing task plans, selecting tools, summarizing system states, and supporting operators during unexpected events. However, they should remain verifiable decision-support components rather than unconstrained controllers. Key research issues include resource-aware model decomposition, model and knowledge freshness, uncertainty estimation, provenance-aware generation, secure tool invocation, and validation against physical observations and mission constraints.

%% file: sections/07_conclusion.tex
\section{Conclusions}

This article introduced StarMind as a distributed orbital computing fabric supported by a five-layer architecture and six standardizable interfaces. The SAR/InSAR case study showed that its benefit depends on workload size, data reduction, OISL capacity, ground contact delay, and resource availability. Rather than replacing terrestrial processing, StarMind should form part of a space–ground computing continuum that dynamically places tasks and data according to E2E service requirements.
